\documentstyle[11pt,paspconf,epsf]{article}

\def\lesssim{\mathrel{\hbox{\rlap{\hbox{\lower4pt\hbox{$\sim$}}}\hbox{$<$}}}}
\def\gtrsim{\mathrel{\hbox{\rlap{\hbox{\lower4pt\hbox{$\sim$}}}\hbox{$>$}}}}

\def\arcdeg{\hbox{$^\circ$}}

\newcommand{\mm}[1]{\mbox{$#1$}}
\newcommand{\unit}[1]{\ifmmode \:\mbox{\rm #1}\else \mbox{#1}\fi}
\renewcommand{\sb}[1]{\mbox{$_{\rm #1}$}}

\newcommand{\mone}{\mm{^{-1}}}

\newcommand{\kms}{\unit{km~s\mone}}

\newcommand{\mpc}{\unit{Mpc}}

\newcommand{\hmpc}{\mm{h\mone}\mpc}

\newcommand{\lb}[2]{\mm{l = #1\arcdeg}, \mm{b = #2\arcdeg}}

\newcommand{\eqref}[1]{equation~(\ref{eq:#1})}

\newcommand{\old}[1]{}

\begin{document}

\title{Are Recent Peculiar Velocity Surveys Consistent?}

\author{Michael J. Hudson\altaffilmark{1}, Russell
J. Smith\altaffilmark{2}, John R. Lucey\altaffilmark{2}, David
J. Schlegel\altaffilmark{3} and Roger L. Davies\altaffilmark{2} }

\altaffiltext{1}{Dept. of Physics \& Astronomy, Univ. of Victoria,
P.O. Box 3055, Victoria, B.C. V8W 3P6, Canada.}
\altaffiltext{2}{Dept. of Physics, Univ. of Durham,
Science Laboratories, Durham, DH1 3LE, United Kingdom.}
\altaffiltext{3}{Dept. of Astrophysical Sciences, Princeton
University, Peyton Hall, Princeton NJ, USA.}

\begin{abstract}
We compare the bulk flow of the SMAC sample to the predictions of
popular cosmological models and to other recent large-scale peculiar
velocity surveys. Both analyses account for aliasing of small-scale
power due to the sparse and non-uniform sampling of the surveys.  We
conclude that the SMAC bulk flow is in marginal conflict with flat
COBE-normalized $\Lambda$CDM models which fit the cluster abundance
constraint.  However, power spectra which are steeper shortward of the
peak are consistent with all of the above constraints. When recent
large-scale peculiar velocity surveys are compared, we conclude that
all measured bulk flows (with the possible exception of that of Lauer
\& Postman) are consistent with each other given the errors, provided
the latter allow for ``cosmic covariance''.  A rough estimate of the
mean bulk flow of all surveys (except Lauer \& Postman) is $\sim
400\kms$ towards \lb{270}{0}.
\end{abstract}


\section{Introduction}

The SMAC cluster sample (see Smith et al., this volume; Hudson et
al. 1999) has a peculiar velocity of $\sim 600 \kms$, with respect to
the Cosmic Microwave Background (CMB) frame, within a depth of $\sim
12000 \kms$.  Other surveys (Willick 1999a,b, also this volume,
hereafter LP10k; Lauer \& Postman 1994, hereafter ACIF) have also
yielded large bulk motions on similarly large scales.  Taken at face
value, these results appear to be in gross conflict with cosmological
models.  However, at the same time, other surveys (notably Dale et
al. 1999, also Dale \& Giovanelli this volume, hereafter SC) have
found rather small bulk motions on similar scales.  Because all of
these surveys are quite sparse, small-scale (``internal'') flows will
not completely cancel, and will act as an extra source of noise.  In
order to allow for these ``aliasing'' effects, it is necessary to
account for the sparse spatial sampling and to have some idea of the
expected level of the internal flows. The latter can be obtained if
the power spectrum of mass fluctuations is known.  To calculate these
effects we will follow the methods of Kaiser (1988) and of Watkins \&
Feldman (1995).  The purpose of this contribution is to address two
questions.  First, what bulk flow do we expect for the SMAC sample,
given currently popular cosmological modes.  Is the SMAC result
consistent with these expectations or does it demand substantial
revision of the models?  Second, are the various large-scale survey
bulk flow statistics consistent with each other, given this extra
small-scale noise?

\section{SMAC vs. cosmological models}

The SMAC sample consists of 56 clusters to a depth of $\sim 12000
\kms$, and so is a rather sparse sample.  The SMAC bulk flow of $630
\pm 200 \kms$ in the CMB frame is inconsistent with zero at the 99.9\%
confidence level (CL).  A sphere of radius 12000 \kms\ would be
expected to have a typical (rms) bulk flow of $\sim 150 \kms$ for a
COBE-normalized $\Lambda$CDM model.  Thus, naively, the SMAC bulk flow
appears to be in gross conflict with the theoretical predictions.  In
order to compare this result with the predictions of cosmological
models, however, the SMAC sample should {\em not} be modeled as a
top-hat sphere.  As first emphasized by Kaiser (1988), it is necessary
to calculate the window function for the survey and multiply this by
the power spectrum to obtain the expected cosmic variance.

\begin{figure}
\plotone{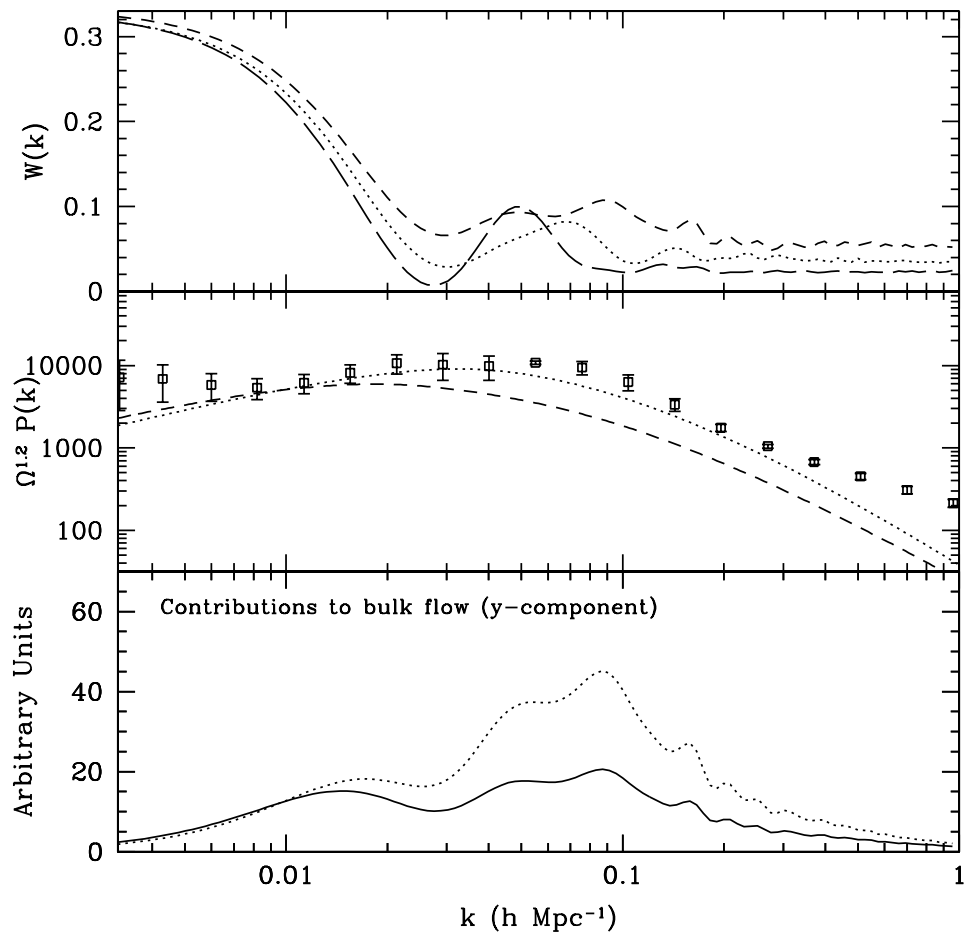}
\caption{Top panel: Window functions for the three Galactic Cartesian
coordinates. x -- dotted; y -- short dash; z -- long dash.  Middle
panel: Power spectra for two models: $\Lambda$CDM (dashed); CHDM
(dotted).  See Table 2 for details of these models. Data points are
from the APM galaxy survey (Gaztanaga \& Baugh 1998). Bottom panel:
the window function of the bulk flow multiplied by the power spectra
($\Lambda$CDM -- dashed; CHDM -- dotted).  This panel shows the wide
range of scales ($0.007 \lesssim k \lesssim 0.2$) that contribute to
the cosmic variance of the SMAC bulk flow statistic.}
\label{fig:win}
\end{figure}

We have calculated the window functions for each Cartesian component
SMAC bulk flow following Kaiser (1988). These are plotted in the top
panel of Fig.\ 1.  Note that the window functions ``ring'' on small
scales ($k > 0.05$).  When the window function is multiplied by the
power spectrum (middle panel), we obtain the contributions to the bulk
flow cosmic variance as a function of scale (bottom panel).  The
contributions to the bulk flow statistic come from a wide range of
scales, with significant contributions from scales as small as
$\lambda \sim 30 \hmpc\ (k \sim 0.2)$.

To assess the consistency with cosmological models, we compute a total
covariance matrix ${\bf C} = {\bf C}\sb{cosmic} + {\bf C}\sb{pv}$,
where the subscript ``cosmic'' denotes the cosmic variance part and
``pv'' denotes the peculiar velocity errors.  For the observed SMAC
bulk flow, we can then compute $\chi^2 = {\bf V}^T \cdot {\bf C}^{-1}
\cdot {\bf V}$, where ${\bf V}$ is the observed bulk flow vector.  By
comparing this statistic with the probability distribution for
$\chi^2$ with three degrees of freedom (corresponding to the three
components of the bulk flow vector) we obtain, for a given
cosmological model, the probability of observing a flow as large as we
do.  As the amount of fluctuation power decreases, so does the ${\bf
C}\sb{cosmic}$, the cosmic variance in the bulk flow, and so $\chi^2$
increases.  This can be used to place constraints on power spectra. For
example, consider the Galactic y-component of the bulk flow, which
dominates the SMAC signal. For the $\Lambda$CDM model of Table 2, the
expected rms cosmic value is $\sim 175$ \kms.  This is considerably
larger than the cosmic rms of $\sim 85$ \kms\ which would be expected
if the sphere to 12000 \kms\ was fully sampled, but is still quite
small compared to the observed bulk flow component of 680 \kms\ in
that direction.  We conclude that this model is excluded at the 97\%
level. On the other hand, if we consider a CHDM model (see Table 2),
the expected cosmic variance increases to 225 \kms, because of the
additional power on intermediate scales (see the middle panel of
Fig. 1). This model is excluded at only the 91\% CL.

In a similar fashion, we can vary the parameters of a given family of
models and ask which combinations yield cosmic variances which are too
small compared to the observed SMAC bulk flow.  In Fig.\ 2, we show
the excluded regions of $\Omega$-$h$ parameter space for
COBE-normalized flat $\Lambda$CDM models with $\Omega\sb{b} = 0.02
h^{-2 }$.  As can be seen from Fig 2., the excluded regions are well
delineated according to the combination $\Omega^{(0.53-0.13\Omega)}
\sigma_8$, where $\sigma_8$ is the rms mass fluctuation in an 8 \hmpc\
sphere.  This same combination also determines the abundance of rich
clusters. The SMAC result require that this combination be $> 0.64$ at
the 95\% level.  This is formally inconsistent with cluster constraint
$\sim 0.55$ (e.g.\ Eke et al.\ 1996), but is consistent with
determinations of $\Omega^{0.6} \sigma_8$ from other peculiar velocity
surveys (Zaroubi et al. 1997; see also Zehavi et al. in this
volume). Note that, if on the other hand we adopt a power spectrum
which is steeper than $\Lambda$CDM on scales smaller than the peak
(e.g.\ CHDM) we do find consistency between the SMAC result and the
cluster constraint.

\begin{figure}
\plotone{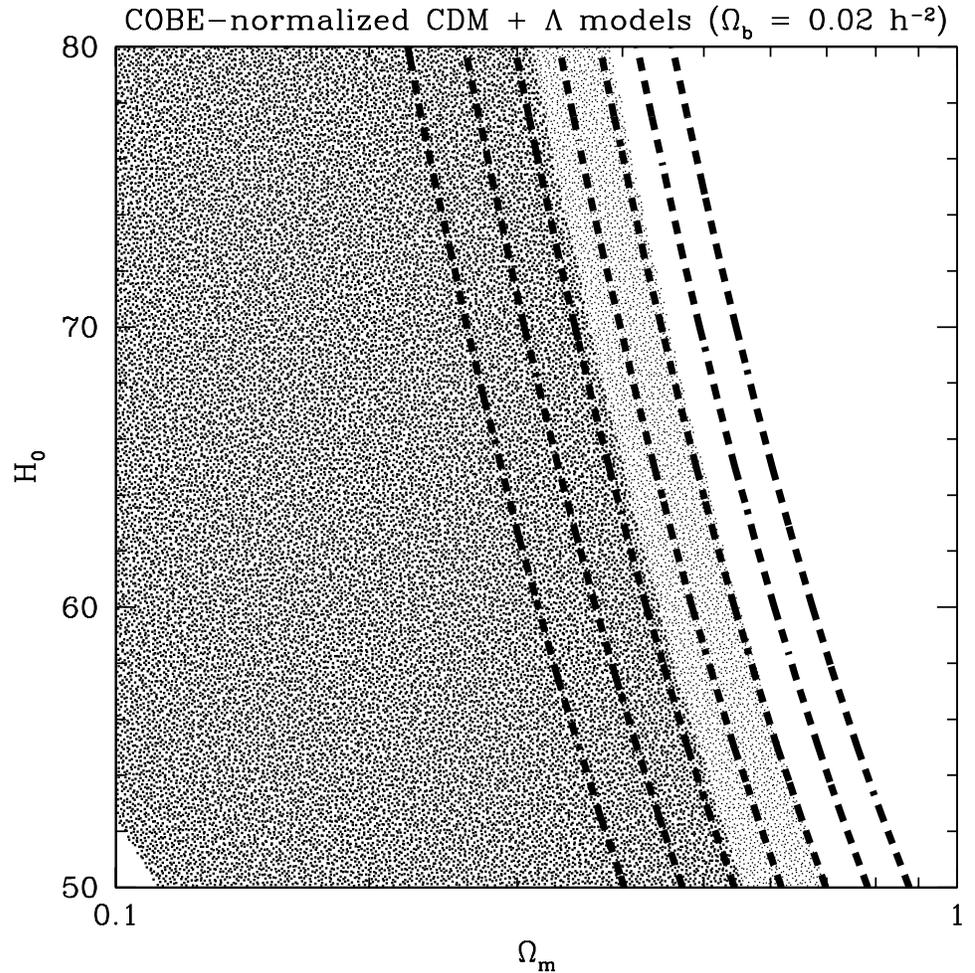}
\caption{Constraints on $\Omega$ and $H_0$ for flat COBE-normalized
$\Lambda$CDM models. The dark grey region indicates the parameter
space excluded at better than the 95\% CL by the observed SMAC bulk
flow. The light grey region shows marginally excluded parameter space
(between 90\% and 95\% CL). The dashed lines indicate combinations of
$\Omega^{(0.53-0.13\Omega)}\sigma_8$ from 0.4 to 1.0 (left to right)
in steps of 0.1.  The cluster abundance constraint yields 0.52 (Eke et
al.\ 1996) for this value, which is excluded at just better than 95\%
by the SMAC bulk flow.  Note that this is a model-dependent result:
for spectra which are steeper on scales shortward of the peak (e.g.\
CHDM) there is overlap between the cluster abundance constraint and
the SMAC bulk flow.}
\label{fig:lambda}
\end{figure}

This shows that, even when allowing for the aliasing effects of sparse
sampling, the bulk motion of the SMAC survey is still rather too large
to be comfortably accommodated by the family of $\Lambda$CDM models.
Therefore, we conclude that there is evidence for excess power on
scales $\sim 30 - 1200 \hmpc$ although, at present, it is significant at the
$\sim 2 \sigma$ level.

\section{Consistency of SMAC, SC, LP10k, ACIF and SNIa}

In this section, we consider results from 5 surveys. In addition to
SMAC, these are: SC, LP10k, ACIF and SNIa (Reiss et al. 1995). We have
measured bulk flows for each of these surveys in a consistent way,
adopting the authors' peculiar velocities and errors, but adding in
quadrature an addition `thermal' scatter of $250 \kms$, to represent
the scatter of individual clusters around the large-scale bulk
flow. This reduces the weight of some nearby well-observed clusters
such as Centaurus. For the ACIF survey, we estimate bulk flows and
errors following Hudson \& Ebeling (1997).  The bulk flow results are
given in Table 1.

\begin{table}
\caption{Bulk flows for different surveys}\label{tab:bf}
\begin{center}
\begin{tabular}{llrrrrr}
\hline 
Survey & 
Method & 
N & 
Depth & 
\multicolumn{1}{c}{$\bf V$} & 
$l$ & 
$b$ \\
\hline 
SMAC & FP & 56 & 6600 & $630\pm200$ & 260 & -1  \\
LP10k & TF & 15 & 11100 & $1000 \pm 438$ & 277 & 27 \\
SC & TF & 63 & 8100 & $104 \pm 119$ & 300 & 18 \\ 
SNIa & SNIa & 24 & 4000 & $444 \pm 194$ & 276 & -8 \\
ACIF & BCG & 119 & 8400 & $832 \pm 252$ & 349 & 51 \\
\hline 
\end{tabular}
\end{center}
\end{table}

The window functions of each survey are quite different, particularly
on small scales.  This is a reflection of the fact that all surveys
react to the same large-scale structures, but the aliased contribution
arising from small scales differs from one survey to another depending
on the spatial sampling.  We therefore expect the measured bulk flows
to be correlated, but not identical, even in the absence of peculiar
velocity errors.  We refer to this as ``cosmic covariance''.  Given a
power spectrum, this cosmic covariance matrix can be quantified,
following the method of Watkins \& Feldman (1995).  Specifically, we
compare with zero the measured difference between the bulk flows of
two surveys A and B, ${\bf V}\sb{A} - {\bf V}\sb{B}$.  In order to
determine whether the observed difference is significant, the error
analysis includes both the peculiar velocity errors and ``cosmic
covariance''.  The latter is calculated by computing the full
covariance matrix of the bulk flow components for the two surveys.
This cosmic covariance term is not negligible. For most comparisons
here, the expected rms difference between bulk flows in the absence of
peculiar velocity errors is still $\sim 200$--300 \kms.

In Table 2, we present a selection of comparisons between pairs of
surveys (e.g.\ SMAC vs. ACIF and SMAC vs. SC), as well as comparisons
of the type Survey A vs. ``All-surveys-except A''.  (The SNIa results
are omitted from this table because they are consistent with all
results). The table lists the probability that two surveys are
consistent with the same underlying peculiar velocity field. This is
done for two representative cosmological models with somewhat
different shaped power spectra. If the ``cosmic covariance'' is
neglected, one would conclude that at least two surveys (ACIF and SC)
are inconsistent with the rest.  However, once cosmic variance is
included, there is no significant conflict between SC and the other
surveys.  The only survey which stands apart is the ACIF survey, and
even then the difference is only marginal (significant at the 93\%
level).\footnote{It is expected that if the EFAR results (Colless et
al., this volume) were included in this analysis, the consistency of
SC would improve but that of ACIF would become worse.}

\begin{table}
\caption{Consistency of Surveys. The table shows the probability that
two surveys are consistent, with the same true velocity field, given
their peculiar velocity errors ('None') or peculiar velocity errors {\em plus}
cosmic covariance, assuming either $\Lambda$CDM or CHDM. Comparisons
which show disagreement at greater than the 95\% CL are in
italics.}\label{tab:consist}
\begin{center}
\begin{tabular}{l@{ vs. }lrrr}
\multicolumn{2}{l}{Surveys} & & \multicolumn{2}{c}{`Cosmic Covariance'} \\
\multicolumn{2}{c}{}& None & $\Lambda$CDM$^{a}$ & CHDM$^{b}$ \\
\hline
\multicolumn{5}{c}{}\\
SMAC & ACIF	& {\em 0.020\/} & {\em 0.022\/} & {\em 0.027\/} \\
SMAC & SC	& {\em 0.025\/} & 0.051 & 0.119 \\
\multicolumn{5}{c}{}\\
\hline
\multicolumn{5}{c}{}\\
ACIF & SMAC+SC+LP10k+SNIa & 0.058 & 0.062 & 0.068 \\
SMAC & SC+LP10k+SNIa	& 0.630 & 0.657 & 0.704\\ 
SC & SMAC+LP10k+SNIa	& {\em 0.033\/} & 0.075 & 0.171 \\
LP10k & SMAC+SC+SNIa	& 0.396 & 0.428 & 0.482 \\
\multicolumn{5}{c}{}\\
\hline
\end{tabular}
\end{center}
\tablenotetext{a}{$\Lambda$CDM: $\Omega_m = 0.35$, $H_0 = 65$, $\Omega_b = 0.047$}
\tablenotetext{b}{CHDM: $N_{\nu}=2$, $\Omega_{\nu}=0.2$, $H_0 = 50$, $\Omega_b = 0.075$}
\end{table}

\begin{figure}
\plotone{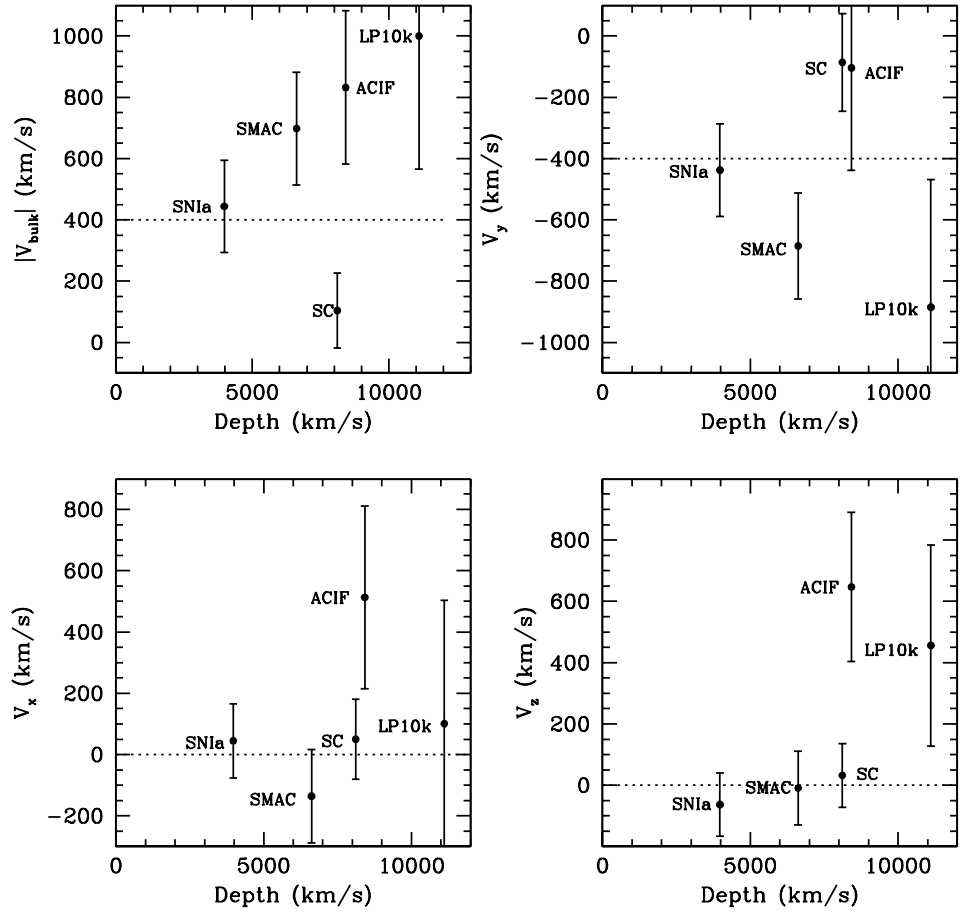}
\caption{Bulk flow amplitude and components in Galactic Cartesian
coordinates for the large-scale peculiar velocity surveys discussed in
the text. The error bars include the usual peculiar velocity errors,
plus an estimate of the aliased small-scale power. The dotted line
shows an eyeball estimate of the mean bulk motion of all surveys
(except ACIF): $\sim 400\kms$ towards Galactic y, \lb{270}{0}.  All
surveys except ACIF are consistent with this bulk motion at better
than the 90\% level.  }\label{fig:bulk}
\end{figure}

Fig 3.\ shows a comparison of the bulk flow components obtained by
different surveys. Here the error bar includes the random errors plus
the contributions of small-scale aliasing (assuming the $\Lambda$CDM
model of Table 2).  The dotted line shows an eyeball estimate of the
mean bulk motion: $\sim 400 \kms$ towards Galactic y, \lb{270}{0}.
All surveys except ACIF are consistent with this bulk motion at better
than the 90\% level. 

\section{Summary}
We have compared the bulk flow of the SMAC sample to the predictions
of popular cosmological models and to other recent large-scale
peculiar velocity surveys. Both analyses account for aliasing of
small-scale power due to the sparse and non-uniform sampling of the
surveys.  We conclude that the SMAC bulk flow is in marginal conflict
with flat COBE-normalized $\Lambda$CDM models which fit the cluster
abundance constraint.  However, power spectra which are steeper
shortward of the peak are consistent with all of the above
constraints. When SMAC is compared to other recent peculiar velocity
surveys, we conclude that all measured bulk flows (with the possible
exception of ACIF) are consistent with each other
given the errors, provided the latter allow for the aliasing of
small-scale power. A rough estimate of the mean bulk flow of all
surveys (except ACIF) is $\sim 400\kms$ towards \lb{270}{0}.



\begin{references}

\reference Dale, D. A., Giovanelli, R. , Haynes, M. P., Campusano, L. E., 
Hardy, E.  \& Borgani, S.  1999, \apjl, 510, L11 (SC)

\reference Eke, V. R., Cole, S., \& Frenk, C. S. 1996, \mnras, 282, 263

\reference Gaztanaga, E. \& Baugh, C. M. 1998, \mnras, 294, 229

\reference Hudson, M. J. \& Ebeling, H.  1997, \apj, 479, 621 

\reference Hudson, M. J., Smith, R. J., Lucey, J. R., Schlegel, D. J. \& 
Davies, R. L. 1999, \apjl, 512, L79  (SMAC)

\reference Kaiser, N. 1988, \mnras, 231, 149

\reference Lauer, T. R. \& Postman, M. 1994,
\apj, 425, 418 (ACIF)

\reference Riess, A. G., Davis, M., Baker, J., \& Kirshner,
R. P., 1997, \apjl, 488, L1 (SNIa)

\reference Watkins, R. \& Feldman, H. A. 1995, \apjl, 453, L73

\reference Willick, J. A. 1999a, \apj, 516, 47 (LP10k)

\reference Willick, J. A. 1999b, submitted to ApJ (astro-ph/9812470)

\reference Zaroubi, S. , Zehavi, I. , Dekel, A. , Hoffman, Y.  \&
Kolatt, T.  1997, \apj, 486, 21

\end{references}
\end{document}